\documentclass[twocolumn,showpacs,preprintnumbers,floats,amsmath,amssymb]{revtex4-1}
\usepackage{graphicx}
\usepackage{dcolumn}
\usepackage{bm}
\usepackage{amsmath}
\usepackage{amsfonts}
\usepackage{amssymb}%
\providecommand{\U}[1]{\protect\rule{.1in}{.1in}}
\begin{document}
\title{The Hubble IR cutoff in holographic ellipsoidal cosmologies}
\author{Mauricio Cataldo}
\altaffiliation{mcataldo@ubiobio.cl}

\affiliation{Departamento de F\'\i sica, Universidad del B\'\i o--B\'\i o, Avenida Collao
1202, Casilla 5-C, Concepci\'on, Chile.\\}

\author{Norman Cruz}
\altaffiliation{norman.cruz@usach.cl}

\affiliation{Departamento de F\'{\i}sica, Universidad de Santiago,}
\affiliation{Casilla 307, Santiago, Chile. \\}
\date{\today}

\begin{abstract}
It is well known that for spatially flat FRW cosmologies, the
holographic dark energy disfavours the Hubble parameter as a
candidate for the IR cutoff. For overcoming this problem, we explore
the use of this cutoff in holographic ellipsoidal cosmological
models, and derive the general ellipsoidal metric induced by a such
holographic energy density. Despite the drawbacks that this cutoff
presents in homogeneous and isotropic universes, based on this
general metric, we developed a suitable ellipsoidal holographic
cosmological model, filled with a dark matter and a dark energy
components. At late time stages, the cosmic evolution is dominated
by a holographic anisotropic dark energy with barotropic equations
of state. The cosmologies expand in all directions in accelerated
manner. Since the ellipsoidal cosmologies given here are not
asymptotically FRW, the deviation from homogeneity and isotropy of
the universe on large cosmological scales remains constant during
all cosmic evolution. This feature allows studied holographic
ellipsoidal cosmologies to be ruled by an equation of state
$\omega=p/\rho$, whose range belongs to quintessence or even phantom
matter.

\vspace{0.5cm}
\end{abstract}

\maketitle

\section{Introduction}
In the description of the early universe, spatially homogeneous and
anisotropic cosmological models may be allowed. These models via
some mechanism, such for example dissipation~\cite{Gron}, could
evolve to an homogeneous and isotropic one. The evidence comes from
the existence of small anisotropy deviations from isotropy of the
CMB radiation and the presence of large angle anomalies, which
represent real features of the CMB map of the
Universe~\cite{Planck}. These anomalies seem to indicate a preferred
orientation in the space, and it is unclear whether they originate
from some unknown systematic error (present in both the COBE and
WMAP data) or if they have a physical origin~\cite{Rodrigues}.

Using a Bianchi type I metric, in~\cite{Russell} it was obtained a
model which becomes to be an almost FRW in time that is consistent
with current data of the CMB. In this work it was assumed that the
matter component forms the deviations from isotropy in the CMB
density fluctuations when matter and radiation decouples. Some
authors suggest that the ellipsoidal cosmological model is a viable
alternative that could account for the detected large scale
anomalies in the cosmic microwave anisotropies~\cite{Cea}, although
the description of polarization modes, specifically B modes, are not
properly described in the framework Bianchi type I
cosmologies~\cite{Pontzen}.

On the other hand, the anisotropy of the universe can be associated
with dark energy, since anisotropic stresses at the perturbative
level are characteristics of various cosmological models of dark
energy, which are compatible with the homogeneity and isotropy of
the FRW geometry~\cite{Koivistovarios}. An explicit field theory for
the anisotropically stressed dark energy in a universe described by
the Bianchi type I metric was formulated in~\cite{Koivisto}, and the
parameters were constrained using the luminosity-redshift
relationship of the SNIa data. For an ellipsoidal universe, which is
Bianchi Type I cosmological model with highest symmetry in the
spacial sections of the spacetime geometry, the actual skewness and
shear of the dark energy component were constrained using Union2
data for supernovae~\cite{Campanelli}. The EoS for dark energy
described by an energy density $\rho_{DE}$ was assumed in this work
of the form $p_{\parallel}=\omega_{\parallel}\rho_{DE}$,
$p_{\perp}=\omega_{\perp}\rho_{DE}$, with $\omega_{\parallel}$ and
$\omega_{\perp}$ constants.

In more formal studies, Bianchi type-I anisotropic cosmological
models have been extensively investigated for a wide types of matter
content. In terms of discussing properties of the dark energy it is
of interest the inclusion of a nonzero cosmological constant in this
type of models.  A detailed analysis of the dynamical systems
corresponding to a Bianchi type-I anisotropic universe filled with a
cosmological constant and a fluid with bulk viscosity was realized
in~\cite{Kohli}. Anisotropic universes of this type filled with
perfect fluid matter with or without dissipative process and a
cosmological constant has been investigated in~\cite{Varios}. A
variable cosmological constant has also been taken into
consideration in related research. This is the case of a magnetized
Bianchi I universe that was investigated in~\cite{Pradhan}. The
inclusion of a bulk viscous fluid was considered
in~\cite{Belinchon}.

Anisotropic dark energy also has been investigated in the framework
of the holographic principle~\cite{Gonzalez}, which is believed to
be a fundamental principle for the quantum theory of gravity. Based
in this principle, holographic dark energy models have been recently
advanced~\cite{Cohen,Hsu,Li}. Therefore these models incorporate
significant features of the underlying theory of dark energy. The
holographic principle is a conjecture stating that all the
information stored within some volume can be described by the
physics at the boundary of the volume and, in the cosmological
context, this principle will set an upper bound on the entropy of
the universe. With the Bekenstein bound in mind, it seems to make
sense to require that for an effective quantum field theory in a box
of size L with a short distance cutoff (UV cutoff: $\Lambda$), the
total entropy should satisfy the relation
\begin{equation}
L^{3}\Lambda^{3}\leq S_{BH} = \pi L^{2}M_{p}^{2},
\label{Bekenstein}
\end{equation}
where $M_{p}$ is the reduced Planck mass and $S_{BH}$ is the
entropy of a black hole of radius $L$ which acts as a long
distance cutoff (IR cutoff: L). However, based on the validity of
effective quantum field theory Cohen et al~\cite{Cohen} suggested
a more stringent bound, requiring that the total energy in a
region of size $L$ should not exceed the mass of a black hole of
the same size. Therefore, this UV-IR relationship gives an upper
bound on the zero point energy density
\begin{equation}\label{zeropoint}
\rho_{\Lambda}\leq L^{-2}M_{p}^{2},
\end{equation}
which means that the maximum entropy is $ S_{max}\approx S^{3/4}_{
BH}$. The largest $L$ is chosen by saturating the bound in
Eq.~(\ref{zeropoint}) so that we obtain the holographic dark energy
density
\begin{equation}
\rho_{\Lambda}=3c^{2}M_{p}^{2}L^{-2},\label{holobound}
\end{equation}
where c is a free dimensionless $\mathcal{O}$(1) parameter and the
coefficient 3 is chosen for convenience. Interestingly, this
$\rho_{\Lambda}$ is comparable to the observed dark energy density
$10^{-10}eV^{4}$ for $H = H_{0}\sim  10^{-33}eV$ , the Hubble
parameter at the present epoch. This means that if we choose the IR
cutoff as the current horizon size we obtain the current observed
dark energy scale. The fact that quantum field theory over-counts
the independent physical degrees of freedom inside the volume
explains the success of this estimate over the value
$\rho_{\Lambda}= \mathcal{O}(M_{p}^{4})$. Therefore, holographic
dark energy models have the advantage over other models of dark
energy in that they do not need an adhoc mechanism to cancel the
$\mathcal{O}(M_{p}^{4})$ zero point energy of the vacuum.

Nevertheless, as it was pointed out by Hsu~\cite{Hsu}, the current
Hubble horizon as IR cutoff in the Friedmann equation $\rho =
3M_{P}^{2}H^{2}$ makes the dark energy behaves like matter rather
than a negative pressure fluid, and prohibits accelerating expansion
of the universe. In fact, in this case we have that $\rho_m
\thicksim H^2$ and $\rho_{_{DE}} \thicksim H^2$. This tracker
behavior of dark components implies that the dark matter and
holographic dark energy scale with the universe scale factor as
$a^{-3}$, leading to a pressureless dark energy.

Due to the above limitation of taking current Hubble horizon as IR
cutoff, other cutoff has been investigated in the framework of an
homogeneous and isotropic cosmology, such as the Ricci scalar
~\cite{Cai} associated to the causal conection scale for
perturbations, the event horizon~\cite{Li}, and the proposed
in~\cite{Granda}, which is of the form $\rho \approx \alpha
H^{2}+\beta \dot{H}^{2}$, where $\alpha, \beta$ are constants. In
the holographic framework, Bianchi Type I has been analyzed for a
universe filled with matter and generalized holographic or
generalized Ricci dark energy, using the statefinder
parameters~\cite{Sharif}. Exact solutions for a homogeneous axially
symmetric Bianchi type I universe filled with matter and holographic
dark energy were found in~\cite{Sarkar}. In this work it was used
the cutoff proposed in~\cite{Granda} and a constant deceleration
parameter was assumed.

The main aim of this paper consists of studying Bianchi type I
cosmologies filled with a holographic dark energy by choosing the IR
cutoff as the size of our universe. For doing this we derive the
general ellipsoidal metric induced by a holographic energy density
of the form~(\ref{holobound}), when $L=H^{-1}$. It is remarkable
that the generated metric allows to consider accelerated expansion
in all directions, which is in agreement with observations. This
behavior is not typical for all Bianchi type I cosmologies since
often there are solutions where simultaneously some directional
scale factors expand while others contract. Another aspect that
deserves consideration is that the obtained holographic metric is
not asymptotically FRW since it is always anisotropic due to the
presence of a constant parameter, which can be constrained by
observations.

This holographic ellipsoidal metric is coupled to compatible matter
sources. We consider accelerating cosmological models filled with an
isotropic dark matter component and an anisotropic holographic dark
energy, satisfying the relations $\rho_m \thicksim H^2$ and
$\rho_{_{DE}} \thicksim H^2$ during all cosmic evolution (or at late
times), where now $H$ is the mean Hubble parameter (see
Eq.~(\ref{mean H})).

Ellipsoidal metrics are homogeneous and anisotropic Bianchi type I
models with the highest (planar) symmetry in the spatial sections of
the geometry. As we stated above, we consider the Hubble length as
the IR cutoff. Despite the drawbacks with this cutoff in obtaining a
well behaved EoS for the dark energy FRW cosmologies, we explore
their properties and consequences in anisotropic metrics, and
constraint them by using the values obtained for the present level
of anisotropy of the large scale geometry of the Universe.

The organization of the paper is as follows: in Sec. II we present
the field equations for a spatially homogeneous and anisotropic
Bianchi type I universe with planar symmetry, and derive the general
ellipsoidal metric induced by the considered holographic energy
density. In Sec. III, we discuss an ellipsoidal cosmological
solution filled with an isotropic dark matter component and an
anisotropic holographic dark energy. This model, as well as the
model with the holographic dark energy dominating the expansion, are
constrained using the data for the actual shear and skewness of the
universe. Finally, in Sec. IV we present the conclusion of our
results.

\section{Anisotropic holographic model and Einstein Field Equations}
We shall consider a particular case of spatially homogeneous and
anisotropic Bianchi type I models described by the line element
\begin{eqnarray}\label{metric}
ds^2=dt^2-a^2(t) (dx^2+dy^2)-b^2(t) dz^2,
\end{eqnarray}
where $a(t)$ and $b(t)$ are the directional scale factors and are
functions of the cosmic time t. This spacetime possesses spatial
sections with planar symmetry, with axis of symmetry directed along
the $z$-axis. The metric~(\ref{metric}) describes a space that has
an ellipsoidal rate of expansion at any moment of the cosmological
time.

In this case the Einstein field equations are given by
\begin{eqnarray}
\kappa \rho=\frac{\dot{a}^2}{a^2}+2 \frac{\dot{a} \, \dot{b}}{a \,
b}, \label{00}\\
\kappa p_1=-\left( \frac{\ddot{a}}{a} + \frac{\dot{a} \, \dot{b}}{a
\, b} +\frac{\ddot{b}}{b} \right), \label{11}\\
\kappa p_3= - \left( 2 \frac{\ddot{a}}{a}+\frac{\dot{a}^2}{a^2}
\right), \label{22}
\end{eqnarray}
where $\kappa = 8\pi G = M_{p}^{-2}$. Note that we have put for the
longitudinal and transversal pressures $p_x=p_y=p_1$ and $p_z=p_3$.

Now on we shall consider that the energy density filling this
universe has a holographic character. At this point we assume that
the IR cutoff for anisotropic universes is the mean Hubble parameter
$H$, i.e. $L = H^{-1}$, therefore the holographic energy density
given by Eq.~(\ref{holobound}) becomes
\begin{eqnarray}\label{holodensity}
\kappa \rho_{_H}= 3 c^2 H^2.
\end{eqnarray}
For the metric~(\ref{metric}) we can define the average scale factor
$\bar{a}(t)$ as
\begin{eqnarray}\label{average sf}
\bar{a}(t)=(a^2(t) \, b(t))^{1/3},
\end{eqnarray}
and the mean Hubble parameter takes the form
\begin{eqnarray}\label{mean H}
H=\frac{\dot{\bar{a}}}{\bar{a}}=\frac{1}{3} \left( 2
\frac{\dot{a}}{a}+\frac{\dot{b}}{b} \right),
\end{eqnarray}
obtaining for the holographic energy the relation
\begin{eqnarray}\label{HDE}
\kappa \rho_{_H} =\frac{c^2}{3} \left( 2
\frac{\dot{a}}{a}+\frac{\dot{b}}{b} \right)^2.
\end{eqnarray}

Thus from Eqs.~(\ref{00}) and~(\ref{HDE}) we have the following
differential equation
\begin{eqnarray}
\frac{\dot{a}^2}{a^2}+2 \frac{\dot{a} \, \dot{b}}{a \,
b}=\frac{c^2}{3} \left( 2 \frac{\dot{a}}{a}+\frac{\dot{b}}{b}
\right)^2,
\end{eqnarray}
which implies that the directional scale factors are related by
\begin{eqnarray}\label{a(t)b(t)}
a(t)= b(t)^{\alpha},
\end{eqnarray}
where
\begin{eqnarray}\label{dalpha15}
\alpha=\frac{3-2c^2 \pm 3 \sqrt{1-c^2}}{4c^2-3},
\end{eqnarray}
and, without any loss of generality, the integration constant has
been set equal to 1, since we can rescale the coordinates $x$ and
$y$.

Note that in order to have real values for the parameter $\alpha$
the condition $0 \leq c^2 \leq 1$ must be required. Thus from
Eq.~(\ref{dalpha15}) we obtain that $0 \leq \alpha \leq 1$ for the
minus sign, while for the plus sign we have that $\alpha \geq 1$ for
$\sqrt{3/4}< c \leq 1$, and $\alpha \leq -2$ for $0 \leq c <
\sqrt{3/4}$. Besides, for $c^{2}=3/4$, $\alpha$ becomes infinity for
the plus sign, while $\alpha \rightarrow 1/4$ for the minus sign.

The holographic metric takes the following form:
\begin{eqnarray}\label{holographic metric}
ds^2=dt^2-b(t)^{2\alpha} (dx^2+dy^2)-b(t)^2 dz^2.
\end{eqnarray}
Clearly this metric becomes isotropic for $\alpha=1$, or
equivalently for $c^2=1$. For $0 \leq c^2 < 1$ we can have models
which expands (or contracts) at different rates at different
directions (for $\alpha >0$), or, as well as occur with vacuum
Kasner cosmology, expands (contracts) only along two perpendicular
axes, and contracts (expands) along the $z$-axis (for $\alpha \leq
-2$).

It is interesting to note that the metric~(\ref{holographic metric})
is characterized by the condition that expansion scalar
$\Theta=u^\alpha_{; \alpha}= (2+\alpha) H_a$ is proportional to
shear scalar $\sigma^2=\frac{1}{2} \sigma_{a b}\sigma^{a b}$.


Let us now consider solutions to these spacetimes in terms of the
pressures of the dark energy fluid.

\subsection{Isotropic pressure}
We begin studying the simplest case where the holographic dark
energy has isotropic pressure. For doing this we put $p_1=p_3=p$
into the field equations~(\ref{00})-(\ref{22}), and by taking into
account Eq.~(\ref{a(t)b(t)}), the metric function takes the form
\begin{eqnarray}
b(t)= \left( c_1 t +c_2 \right)^\frac{1}{(2\alpha+1)},
\end{eqnarray}
where $c_1$ and $c_2$ are integration constants. In this case the
energy density and pressure are given by
\begin{eqnarray}\label{stiff pressure}
\rho_{_H}=p=\frac{\alpha(\alpha+2)c_1^2}{\kappa (2 \alpha+1)^2 (c_1
t+ c_2)^2}.
\end{eqnarray}
This means that the isotropic requirement for the pressure implies
that the holographic matter filling the universe is a stiff one, and
the holographic dimensionless parameter may be written through the
relevant model parameter $\alpha$ as
\begin{eqnarray}
c^2 = \frac{3 \alpha(\alpha+2)}{(2\alpha+1)^2}.
\end{eqnarray}
The energy density is positive for $\alpha< -2$ or $\alpha>0$. The
metric of Bianchi Type I in this case of isotropic pressure takes
the form
\begin{eqnarray}\label{holographic isotropic metric}
ds^2=dt^2-(c_1 t+c_2)^{\frac{2 \alpha}{2 \alpha+1}} (dx^2+dy^2)
\nonumber \\ -(c_1 t+c_2)^\frac{2}{2 \alpha+1} dz^2.
\end{eqnarray}
This one-parametric family of anisotropic metrics is the Kasner
metric for a stiff fluid. The scale factor of the symmetric plane
increases as $t^{\alpha/(2 \alpha +1)}$, which means that for
$\alpha > 0$ and $\alpha < -1$ there is no accelerated expansion.
For $-1< \alpha < -1/2$ there is an accelerated expansion of the
symmetric plane and a contraction along the $z$-axis. For $\alpha>0$
there is no accelerated expansion in all directions.

In order to consider more general solutions than those provided by
stiff holographic energy, we can require for the pressures the
following isotropic barotropic equation of state (EoS):
\begin{eqnarray}\label{barot hol 1}
p_1=p_3=\omega \rho_{_H},
\end{eqnarray}
where $\omega$ is a constant state parameter. From
Eqs.~(\ref{00}),~(\ref{11}) and~(\ref{barot hol 1}) we obtain
\begin{eqnarray}\label{b2}
b(t)=(c_1 t +c_2)^{\frac{\alpha+1}{\alpha^2 \omega+\alpha^2+2 \omega
\alpha + \alpha +1}},
\end{eqnarray}
while from Eqs.~(\ref{00}),~(\ref{22}) and~(\ref{barot hol 1}) we
have that
\begin{eqnarray}\label{b1}
b(t)=(c_1 t +c_2)^{\frac{2}{3 \alpha+\omega \alpha +2\omega}}.
\end{eqnarray}

Thus the power-law expressions~(\ref{b2}) and~(\ref{b1}) imply the
following constraint
\begin{eqnarray}\label{constraint isotropo}
\alpha^2 \omega-\alpha^2 +\omega \alpha-\alpha-2\omega+2=0.
\end{eqnarray}
From this relation we obtain that $\omega=1$ for any $\alpha$, or
$\alpha=1,-2$ for any $\omega$. The case $\omega=1$ for any $\alpha$
was discussed before and describes a stiff holographic energy. The
second case $\alpha=1$ for any $\omega$ describes the standard
isotropic FRW models with scale factor given by $a(t)=b(t)=a_0
t^{2/(3 \omega+3)}$. The third case $\alpha=-2$ describes a vacuum
Kasner anisotropic spacetime given by
\begin{eqnarray}\label{metric 2+1}
ds^2=dt^2-t^{4/3} (dx^2+dy^2)-  t^{-2/3} dz^2.
\end{eqnarray}
In conclusion, the only relevant non vacuum solution with
anisotropic pressure is described by the metric~(\ref{holographic
isotropic metric}) and the stiff holographic energy~(\ref{stiff
pressure}). The condition $a(t)=b(t)^{\alpha}$ is fundamental to
obtaining this result. Therefore, it is not possible to describe
accelerated expansion in ellipsoidal cosmologies filled with
isotropic dark energy.

\section{Ellipsoidal universes with anisotropic pressure}
Now we shall consider anisotropic holographic models with
anisotropic pressures $p_1 \neq p_3$. In general the Einstein field
equations for a Bianchi type I metric may be written in the
following form~\cite{Chimento}:
\begin{eqnarray} \label{E1}
3H^2=\kappa \rho+\frac{\sigma^2}{2}, \\ \label{E2}
-2 \dot H=\kappa (\rho + p) +\sigma^2, \\ \label{E3}
\dot \rho+ 3 H (\rho+p) = \vec \sigma \cdot \vec \Gamma, \\
\label{E4}
\dot{ \vec{\sigma}}+3H \vec \sigma=\vec \Gamma,
\end{eqnarray}
where $\kappa=8 \pi G$ (we will consider $\kappa=1$ from here on),
$H$ and $p$ are the average expansion rate and the average pressure.
The new physical quantities $\vec \sigma$ and $\vec \Gamma$ are the
shear vector and the transverse pressure vector respectively, and
are defined as
\begin{eqnarray}\label{B1}
\sigma_i=H_i-H,  \label{B3}  \\
\label{B4} \Gamma_i=p_i-p,
\end{eqnarray}
where $i=1,2,3$. From Eqs.~(\ref{B1})-(\ref{B4}) we see that the
quantities $\vec \sigma$ and $\vec \Gamma$ satisfy the constraints
\begin{eqnarray}
\sigma_1+\sigma_2+\sigma_3=0, \\
\Gamma_1+\Gamma_2+\Gamma_3=0,
\end{eqnarray}
respectively.

From the ellipsoidal metric~(\ref{metric}) we have that
\begin{eqnarray}
\sigma^2=\frac{2}{3} (H_1-H_3)^2, \\
\vec{\sigma}\cdot \vec{\Gamma}=\frac{2}{3} \, (p_3-p_1)(H_3-H_1).
\label{termino de desviacion}
\end{eqnarray}

In the following subsections we shall consider different holographic
models, filled with an isotropic and anisotropic dark components,
and we will contrast them with observational data.

\subsection{Tracker ellipsoidal holographic solution with dark matter and dark energy}
First, we shall study the ellipsoidal version of the tracker FRW
holographic cosmology for which the IR cutoff is the Hubble
parameter~\cite{Hsu,Li}. In order to do this, we shall use the
holographic spacetime~(\ref{holographic metric}) filled with an
isotropic dark matter component and a holographic dark energy with
anisotropic pressures. It becomes clear that if the total energy
includes the dark matter and dark energy density, we can develop a
tracker cosmological model by writing for dark components the
relations
\begin{eqnarray} \label{c1c}
\rho_{_{m}} =3 c_1^2 H^2, \\
\rho_{_{DE}}=3 c_2^2 H^2, \label{c2c}
\end{eqnarray}
then the total energy density is given by
\begin{eqnarray}\label{HDE15}
\rho=\rho_m+\rho_{_{DE}}=3c^2 H^2,
\end{eqnarray}
where $c^2=c_1^2+c_2^2$.

We shall suppose that the dark matter and dark energy are not
interacting and then the isotropic and anisotropic components are
conserved separately. These conditions are imposed by requiring for
the conservation equation~(\ref{E3}) that
\begin{eqnarray}\label{matt}
\dot{\rho}_m &+& 3H \rho_m=0, \\
\dot{\rho}_{_{DE}} &+& 3H(\rho_{_{DE}}+p_{_{DE}}) + 
\frac{2}{3} (p_1-p_3)(H_1-H_3)=0. \nonumber \\ \label{DEDE}
\end{eqnarray}
For the metric~(\ref{holographic metric}) the anisotropic pressures
have the form
\begin{eqnarray}
p_1= -\frac{(\alpha+1)\ddot{b}}{b}-\alpha^2 \frac{\dot{b}^2}{b^2}, \label{p1h} \\
p_3=-\alpha(3\alpha-2) \frac{\dot{b}^2}{b^2}-2 \frac{\ddot{b}}{b}.
\label{p3h}
\end{eqnarray}
Notice that the dark matter is a pressureless perfect fluid, then
$p_1$ and $p_3$ are the pressures of the anisotropic dark energy.

From Eq.~(\ref{matt}) we have that
\begin{eqnarray}\label{rhom}
\rho_m(t)=\rho_{m0} b^{-1-2\alpha},
\end{eqnarray}
where $\rho_{m0}$ is a constant of integration. From
Eqs.~(\ref{DEDE})-~(\ref{p3h}) we have that
$\rho_{_{DE}}(t)=\tilde{C} b^{-1-2\alpha}+\alpha (2+\alpha) \,
\frac{\dot{b}^2}{b^2}$, where $\tilde{C}$ is an integration
constant. The Friedmann equation~(\ref{E1}) imposes that
$\tilde{C}=-\rho_{m0}$, and then we have that the energy density of
the dark component is given by
\begin{eqnarray}\label{rhoDE}
\rho_{_{DE}}(t)=\alpha (2+\alpha) \, \frac{\dot{b}^2}{b^2}-\rho_{m0}
b^{-1-2\alpha}.
\end{eqnarray}
We can find the tracker ellipsoidal version for the considered
holographic cosmology~(\ref{holographic metric}) by imposing on
energy densities the conditions~(\ref{c1c}) and~(\ref{c2c}). In such
a way, from Eqs.~(\ref{rhom}) and~(\ref{c1c}) we obtain that the
scale factor is given by
\begin{eqnarray}\label{bbb}
b(t)=\left( \frac{3 \rho_{m0} \left(t+C\right)^2}{4c_1^2}
\right)^{\frac{1}{2 \alpha+1}},
\end{eqnarray}
where $C$ is a constant of integration. Then we have that
\begin{eqnarray}
\rho_{m}=\frac{4c_1^2}{3(t+C)^2}.
\end{eqnarray}

From Eqs.~(\ref{rhoDE}) and~(\ref{c2c}) we also obtain that $b(t)
\thicksim (t+C)^{2/(2 \alpha+1)}$, but the solution must be self
consistent, then we shall put the scale factor~(\ref{bbb}) into
Eq.~(\ref{rhoDE}), obtaining
\begin{eqnarray}
\rho_{_{DE}}=\frac{4\alpha(\alpha+2)}{(2\alpha+1)^2
(t+C)^2}-\frac{4c_1^2}{3(t+C)^2}.
\end{eqnarray}
For dark energy pressures we obtain
\begin{eqnarray}\label{ppp1}
p_1=\frac{2(\alpha-1)}{(2 \alpha+1)^2(t+C)^2}, \\
p_2=\frac{4(3\alpha-1)(1-\alpha)}{(2 \alpha+1)^2(t+C)^2}.
\label{ppp3}
\end{eqnarray}
Let us now suppose that the longitudinal and transversal pressures
of the dark energy are given by
\begin{eqnarray}\label{pppp15}
p_1=\omega_{1_{DE}} \rho_{_{DE}}, \\
p_3=\omega_{3_{DE}} \rho_{_{DE}},\label{pppp35}
\end{eqnarray}
respectively, where $\omega_{1_{DE}}$ and $\omega_{3_{DE}}$ are
state parameters, which in general are function of the cosmological
time. The pressures $p_1$ and $p_3$ represent the longitudinal and
transversal pressures of the holographic dark energy, since the dark
matter is a pressureless cosmic fluid. In this case the state
parameters are constant and are given by
\begin{eqnarray}\label{omega1de}
\omega_{1_{DE}}=\frac{3(1-\alpha)}{2(-3\alpha(\alpha+2)+c_1^2
(1+2\alpha)^2)}, \\
\omega_{3_{DE}}=\frac{3(1-\alpha)(1-3\alpha)}{-3\alpha(\alpha+2)+c_1^2
(1+2\alpha)^2}.\label{omega3de}
\end{eqnarray}
Note that for $\alpha=1$ we obtain the FRW model, with $b(t)
\thicksim t^{2/3}$ and $p_1=p_2=0 $ (or equivalently
$\omega_{1_{DE}}=\omega_{3_{DE}}=0$), so the holographic energy
density behaves like pressureless fluid as we would expect. For
$\alpha \neq 1$ the pressures $p_1 \neq 0$ and $p_3 \neq 0$, and the
holographic energy becomes anisotropic.

In order to have an accelerated expansion in all directions
Eq.~(\ref{bbb}) and $a(t)=b(t)^\alpha$ imply that
\begin{eqnarray}\label{cond1}
\frac{2}{2\alpha+1}>1,  \\
\frac{2\alpha}{2\alpha+1}>1. \label{cond2}
\end{eqnarray}
It is clear that the $\alpha$-parameter must be positive for having
increasing scale factors. Then, from Eq.~(\ref{cond1}) we have that
$0 <\alpha<1/2$, while condition~(\ref{cond2}) is not possible to
satisfy since $0 \leq \frac{2\alpha}{2\alpha+1} <1$ for $0 \leq
\alpha <\infty$. This implies that we have for $0 < \alpha<1/2$ an
accelerated expansion only in the $x$ and $y$ directions, while in
the $z$-direction the expansion is decelerated.

In conclusion, the tracker ellipsoidal version is mathematically
self-consistent with non vanishing pressures for the holographic
energy, however this solution is ruled out since we have an
accelerated expansion only in two directions: in the third direction
the expansion is decelerated.

\subsection{Ellipsoidal scenarios with dominating holographic dark energy}
Now we shall consider anisotropic scenarios where the holographic
dark energy component dominates over the dark matter content. In
such a way, this model will describes late time stages in the
evolution of an ellipsoidal cosmology where the contribution of the
dark matter density is neglected, and the anisotropic behavior will
be kept due to the presence of anisotropic holographic dark energy
with barotropic anisotropic pressure,
satisfying~(\ref{holodensity}).

Let us suppose that the longitudinal and transversal pressures of
the holographic dark energy are given by
\begin{eqnarray}\label{pp15}
p_1=\omega_1 \rho, \\
p_3=\omega_3 \rho,\label{pp35}
\end{eqnarray}
respectively, where $\omega_1$ and $\omega_3$ are a constant state
parameters (from now on in this section we use the notation
$\omega_{1_{DE}}\equiv \omega_1$ and $\omega_{3_{DE}}\equiv
\omega_3$). Thus, by taking into account that $a(t)=b(t)^\alpha$,
from Eqs.~(\ref{00}) and~(\ref{11}) we obtain that
\begin{eqnarray}\label{bcomun}
b(t)=(c_1 t+c_2)^{\frac{\alpha+1}{\alpha^2 \omega_1
+\alpha^2+2\alpha \omega_1 +\alpha+1}},
\end{eqnarray}
and the metric~(\ref{holographic metric}) takes the following form
\begin{eqnarray}\label{metrica final}
ds^2=dt^2-t^{\frac{\alpha(\alpha+1)}{\alpha^2 \omega_1
+\alpha^2+2\alpha
\omega_1 +\alpha+1}} (dx^2+dy^2) - \nonumber \\
t^{\frac{\alpha+1}{\alpha^2 \omega_1 +\alpha^2+2\alpha \omega_1
+\alpha+1}} dz^2.
\end{eqnarray}
The energy density and the pressure $p_3$ are given by
\begin{eqnarray}
\rho_{_{DE}}=\frac{\alpha (\alpha+2)(\alpha+1)^2}{(\alpha^2 \omega_1
+\alpha^2+2\alpha \omega_1 +\alpha+1)^2 \, t^2}, \label{densidad}
\end{eqnarray}
\begin{eqnarray}\label{pp22}
p_3=\frac{1+2\alpha \omega_1-\alpha}{1+\alpha} \, \rho_{_{DE}},
\end{eqnarray}
respectively. From Eq.~(\ref{pp22}) we conclude that the state
parameter of the transversal pressure is given by
\begin{eqnarray}\label{omega22}
\omega_{3} =\frac{1+2\alpha \omega_1-\alpha}{1+\alpha}.
\end{eqnarray}
Let us now study the deviation of this model from the assumed
homogeneity and isotropy of the universe on large cosmological
scales. From Eqs.~(\ref{omega22}) and~(\ref{delta}) we conclude that
the dark energy skewness parameter takes the form
\begin{eqnarray}\label{deltaFinal}
\delta_{_{DE}}=\frac{(1-\omega_1)(1-\alpha)}{1+\alpha}.
\end{eqnarray}

It is interesting to note that Eq.~(\ref{termino de desviacion}) may
be rewritten with the help of Eqs.~(\ref{shear1})
and~(\ref{deltaFinal}) in the following form
\begin{eqnarray}\label{termino de desviacion reescrito}
\vec{\sigma}\cdot \vec{\Gamma}=2 \delta_{_{DE}} \Sigma H
\rho_{_{DE}}.
\end{eqnarray}
Then the conservation equation for the dark energy
component~(\ref{DEDE}) may be rewritten as
\begin{eqnarray}\label{pdefinal}
\dot{\rho}_{_{DE}}+ 3 H \left(1+ \omega_{eff} + \frac{2}{3} \,
 \delta_{_{DE}} \, \Sigma \right) \rho_{_{DE}}=0,
\end{eqnarray}
where $\omega_{eff}=(2 \omega_1+\omega_{3})/3$. Since the state
parameters, skewness and cosmic shear are constant, then for scaling
scenarios where the dark energy is the dominating component the
quantity $\delta_{_{DE}} \, \Sigma$ is constant, and from
Eq.~(\ref{pdefinal}) we have
\begin{eqnarray}\label{rhobar}
\rho=\rho_0 \, \bar{a}^{-3(1+\omega_{eff})-2 \delta_{_{DE}} \Sigma}.
\end{eqnarray}
Thus, the quantity $\delta_{_{DE}} \Sigma$ characterizes the
deviation from the standard isotropic FRW model, remaining constant
during all evolution of the anisotropic holographic cosmology. Note
that if $\omega_1=1$, then $\omega_{eff}=1$, $\omega_{3}=1$ and
$\delta_{_{DE}}=0$, and we obtain the anisotropic stiff holographic
solution discussed in the previous section. For $\alpha=1$ we have
that $\Sigma=\delta_{_{DE}}=0$, obtaining the standard isotropic FRW
model.

Now we shall assume that the ranges of current shear and skewness
values, obtained in Ref.~\cite{Campanelli}, characterizes the
deviation from the isotropy of a universe dominated, at late times,
by an anisotropic dark energy. Then, we shall use these values to
constraint the parameters of our holographic dark energy model. With
these constraints upon our model we can find its degree of
consistence in terms of the range allowed for the effective EoS of
the holographic dark energy component. As we will show below the
corresponding EoS lies in the range of quintessence or even phantom
dark energy.

However, it is important to note that here we deal with an exact
solution, and it can be shown that $\Sigma$, $\delta_{_{DE}}$ and
$\omega_{eff}$ are not all independent quantities. In general, in
ellipsoidal cosmologies each of these three cosmological parameters
may be written as functions of the scale factors with their
derivatives and constants of integration. In the specific case of
the holographic ellipsoidal cosmology~(\ref{metrica final}), we have
that the relation
\begin{eqnarray}\label{condsigmadelta}
\omega_{eff}=1+\frac{2 \delta_{_{DE}}}{3\Sigma}
\end{eqnarray}
is fulfilled.



As we stated above, Bianchi type I cosmologies are very useful to
test possible anisotropies of the Universe. So, it is interesting to
contrast with observations, the deviation of considered by us
anisotropic models from the assumed homogeneity and isotropy of the
universe on large cosmological scales. In Ref.~\cite{Campanelli} an
ellipsoidal universe is considered, assuming that the mater source
is composed by a noninteracting isotropic pressureless dark matter
and an anisotropic dark energy component. Solving numerically the
Einstein field equations and analyzing the magnitude-redshift data
of type Ia supernovae, it was shown that Supernova data are
compatible with a large level of anisotropy, both in the geometry of
the Universe and in the EoS of dark energy: authors give best-fit
values, and the $1\sigma$ and $2\sigma$ confidence level intervals
derived from the Union2 data analysis, for the cosmologically
relevant parameters $\Sigma$, $\delta$, $w_{eff}$, and $\Omega_m$.

In such a way, for constraining the model parameters, we shall
consider the deviation from isotropy of the EoS, and the amount of
anisotropy in the geometry~(\ref{holographic metric}) by calculating
the cosmic shear $\Sigma$, the skewness $\delta$ and the effective
state parameter $\omega_{eff}$.

Let us introduce the cosmic shear $\Sigma$, defined
by~\cite{Campanelli}
\begin{eqnarray}\label{shear}
\Sigma =\frac{H_{1}-H}{H},
\end{eqnarray}
where $H=\dot{\bar{a}}/\bar{a}$ is the mean Hubble parameter defined
by Eq.~(\ref{mean H}), and $H_{1}= \dot{a}/a$ is the Hubble
parameter for the spatial section of metric~(\ref{metric}). The
parameter $\Sigma$ characterizes the amount of anisotropy in the
geometry since from Eq.~(\ref{shear}) we obtain that $\Sigma \sim
(\dot{a}/a-\dot{b}/b)/H$.

For the holographic metric~(\ref{holographic metric}), the cosmic
shear~(\ref{shear}) takes the form
\begin{eqnarray}\label{shear1}
\Sigma =\frac{\alpha -1}{2\alpha +1}.
\end{eqnarray}
It must be noticed that in general the cosmic shear is time
dependent, however in this case it is constant thanks to the
relation $a(t)=b(t)^\alpha$. With the help of Eqs.~(\ref{dalpha15})
and~(\ref{shear1}) we can constraint the holographic parameter $c$.

The deviation from isotropy of the EoS of the dark energy we shall
characterize with the help of the skewness parameter
$\delta_{_{DE}}$ defined by
\begin{eqnarray}\label{delta}
\delta_{_{DE}}=\omega_{3_{DE}}-\omega_{1_{DE}}.
\end{eqnarray}
It becomes clear that the deviation from isotropy depends only on
the anisotropic character of the dark energy, since the dark matter
fluid is a pressureless one.

From the analysis made in Ref.~\cite{Campanelli} we have that the
present level of anisotropy of the large scale geometry of the
Universe, the actual shear $\Sigma_0$, and the amount of deviation
from isotropy of the EoS of dark energy, the skewness
$\delta_{_{DE}}$, are constrained in the ranges
\begin{eqnarray}\label{sigmaa}
- 0.012 < \Sigma_0 < 0.012,  \\
\label{deltaa} -0.016 \leq \delta_{_{DE}} \leq 0.12,
\end{eqnarray}
respectively.


From Eqs.~(\ref{shear1}) and~(\ref{sigmaa}), and
Eqs.~(\ref{deltaFinal}) and~(\ref{deltaa}), we obtain the
constraints in the form
\begin{eqnarray}\label{sigmaaQ}
- 0.012 < \frac{\alpha -1}{2\alpha +1} < 0.012, \\
- 0.016 < \frac{(1-\omega_1)(1-\alpha)}{1+\alpha} < 0.012,
\label{deltaaQ}
\end{eqnarray}
respectively.

On the other hand, we are interested in describing accelerated stage
of the universe, so we need to request that this model expands in
all directions in an accelerated way, by requiring $\alpha > 0$ and
\begin{eqnarray*}
\frac{\alpha(\alpha+1)}{\alpha^2 \omega_1 +\alpha^2+2\alpha \omega_1
+\alpha+1} > 1, \\
\frac{\alpha+1}{\alpha^2 \omega_1 +\alpha^2+2\alpha \omega_1
+\alpha+1}
> 1.
\end{eqnarray*}
These inequalities imply that
\begin{eqnarray} \label{CC15}
\omega_1 < - \frac{1}{\alpha(\alpha+2)}, \\
\omega_1 < - \frac{\alpha}{\alpha+2}, \label{CC25}
\end{eqnarray}
respectively.

In order to constraint the model parameters we must use the
inequalities~(\ref{ineq DEF}), ~(\ref{sigmaaQ}), ~(\ref{deltaaQ}),
~(\ref{CC15}) and~(\ref{CC25}).

From Eq.~(\ref{sigmaaQ}) we obtain that the parameter $\alpha$ is
constrained as follows:
\begin{eqnarray}\label{constrain alpha}
 0.96484 < \alpha  <  1.03688.
\end{eqnarray}
The constraint on the $\omega_1$-parameter follows from
Eqs.~(\ref{ineq DEF}), (\ref{deltaaQ}), (\ref{CC15})
and~(\ref{CC25}). By taking into account the
constraint~(\ref{constrain alpha}) on the $\alpha$-parameter
Eqs.~(\ref{CC15}) and~(\ref{CC25}) give
\begin{eqnarray}\label{omegaconstarint}
\omega_1< - 0.3496.
\end{eqnarray}
Now, the constraint~(\ref{deltaaQ}) allows $\omega_1$-parameter to
take any value satisfying Eq.~(\ref{omegaconstarint}). Effectively,
we can see that for a given value of $\omega_1$ (even for too big
values $|\omega_1|$) always there exist values for
$\alpha$-parameter, very close to $1$ such that Eq.~(\ref{deltaaQ})
will be satisfied.

Therefore, we have shown that for the considered ellipsoidal
holographic universe~(\ref{metrica final}), filled with an
holographic energy with density~(\ref{holodensity}) and anisotropic
pressures~(\ref{pp15}) and~(\ref{pp35}), the deviation from the
assumed homogeneity and isotropy of the universe on large
cosmological scales remains constant during all evolution of this
type of anisotropic cosmology if the state parameter $\omega_1$
satisfies the constraint~(\ref{omegaconstarint}). Note that
Eqs.~(\ref{constrain alpha}) and~(\ref{omegaconstarint}) imply that
the transversal pressure satisfies the constraint
$\omega_3<-0.3256$, and then in general the holographic energy is
characterized by a quintessence or phantom anisotropic dark energy
EoS.

Now from Eqs.~(\ref{shear1}), (\ref{deltaFinal})
and~(\ref{condsigmadelta}) we find that
\begin{eqnarray}
\omega_{eff}=\frac{2\omega_1-\alpha+1+4\alpha\omega_1}{3(1+\alpha)}.
\end{eqnarray}
Therefore, from Eqs.~(\ref{constrain alpha})
and~(\ref{omegaconstarint}) we conclude that the effective state
parameter has the upper bound $\omega_{eff}<-0.3415$, so the
effective parameter of state may describe a dark energy component
with a negative pressure. To find the lower bound we consider that
in this case the average scale factor is given by $\bar{a}(t)=t^m$,
where
$m=\frac{(2\alpha+1)(\alpha+1)}{3(\alpha^2\omega_1+\alpha^2+2\alpha
\omega_1+\alpha+1)}$. For $\alpha$ in the range~(\ref{constrain
alpha}) the average scale factor describes an accelerated expansion
if $-\sqrt{3/4}<\omega_1<-0.3496$ (in this case $m>1.0122$ and the
expression $\alpha^2\omega_1+\alpha^2+2\alpha \omega_1+\alpha+1$
does not vanish). For $\alpha$ satisfying the
constraint~(\ref{constrain alpha}) and $\omega_1<-\sqrt{3/4}$ the
average scale factor also may describe accelerated expansion. In
this case the expression $\alpha^2\omega_1+\alpha^2+2\alpha
\omega_1+\alpha+1$ vanishes at $\alpha_{\pm}=-\frac{2 \omega_1+1\pm
\sqrt{4 \omega_1^2-3}}{2(1+\omega_1)}$ and we need to study each
case separately. However, despite this, it can be shown that there
exist regimes with $m>1$ for $-1.0122<\omega_1<-\sqrt{\frac{3}{4}}$.

This result imposes the lower bound $-1.0243$ on the effective state
parameter, implying finally that this parameter satisfies the
constraint $-1.0243<\omega_{eff}<-0.3415$, allowing to have
holographic ellipsoidal universes driven by a quintessence or
phantom matter component.

It is interesting to note that CMB data provide tighter constraints
on the anisotropy than the SNeIa data. Specifically, for Bianchi
type I models the present shear is constrained by $\sigma/\theta
\lesssim 10^{-9}$~\cite{Russell}, and since for metric~(\ref{metrica
final}) is valid $\sigma/\theta=\sqrt{\frac{2}{3}} \,
\frac{H_1-H_3}{2H_1+H_3}$, we obtain that $1 \leq \alpha \leq
1.000000004$. Therefore we have for the cosmic shear $0 \leq \Sigma
\leq 1.33333 \times 10^{-9}$ and $\omega_1 \leq \omega_3 \leq
1.000000002 \, \omega_1-2 \times 10^{-9}$. This implies that
$\omega_3 \gtrapprox \omega_1$. Note that from Eqs.~(\ref{CC15})
and~(\ref{CC25}) we obtain that $\omega_1<-1/3$, thus
$\omega_3<-1/3$ (including $\omega_{eff}$), and then the holographic
ellipsoidal model may describe accelerated expansion driven by dark
energy or even phantom matter. This is possible due to the
metric~(\ref{metrica final}) is not asymptotically FRW spacetime.

\subsection{Ellipsoidal cosmology with asymptotic behavior determined by the holographic dark energy}
Now, we are interested in constructing an ellipsoidal cosmological
solution, filled with dark matter and dark energy, whose asymptotic
metric for late times is of the form of Eq.~(\ref{holographic
metric}). In order to do this we shall impose the following
condition on energy densities and longitudinal pressure:
\begin{eqnarray}\label{condprho}
p_1=\omega_1 \left(\rho_m+\rho_{_{DE}} \right),
\end{eqnarray}
where $\omega_1$ is a constant parameter.

By taking into account that $a(t)=b(t)^\alpha$, from
Eqs.~(\ref{00}), (\ref{11}) and~(\ref{condprho}) we obtain that the
directional scale factor $b(t)$ and the ellipsoidal metric are given
by Eqs.~(\ref{bcomun}) and~(\ref{metrica final}), respectively. The
energy densities and transversal pressure in this case are given by
\begin{eqnarray}\label{rhomm}
\rho_m(t)=\rho_{m0} t^{-\frac{\left( \alpha+1 \right)  \left(
2\,\alpha+1 \right)}{{\alpha}^{2}{\it \omega_1}+{\alpha}^{2}+2\,{\it
\omega_1}\,\alpha+\alpha+1}}, \\
\rho_{_{DE}}(t)= \frac {\alpha\, \left( \alpha+1 \right) ^{2} \left(
\alpha+2 \right) }{ \left( {\alpha}^{2}{\it
\omega_1}+{\alpha}^{2}+2\, {\it \omega_1}\,\alpha+\alpha+1 \right)
^{2}{t}^{2}}- \nonumber
\\ \rho_{m0} t^{-{\frac { \left( \alpha+1 \right)  \left( 2\,\alpha+1
\right) }{{\alpha}^{2}{\it
\omega_1}+{\alpha}^{2}+2\,{ \it \omega_1}\,\alpha+\alpha+1}} }, \label{rhodede}\\
p_3(t)=\frac {\alpha\, \left( \alpha+1 \right) \left( \alpha+2
\right) (1+2 \alpha \omega_1-\alpha)}{ \left( {\alpha}^{2}{\it
\omega_1}+{\alpha}^{2}+2\, {\it \omega_1}\,\alpha+\alpha+1 \right)
^{2}{t}^{2}},
\end{eqnarray}
respectively.

It is interesting to note that the dark matter
component~(\ref{rhomm}) satisfies the conservation
equation~(\ref{matt}), while the dark energy~(\ref{rhodede})
satisfies Eq.~(\ref{DEDE}), so they are conserved separately and
there is not change of energy between these dark components.

Notice that in order to find the obtained holographic solution we
have not used Eqs.~(\ref{E2}) and~(\ref{E4}). In this regard, we can
see that by taking into account the metric~(\ref{metric}), and
imposing on Eq.~(\ref{E2}) the holographic
condition~(\ref{holodensity}) we obtain Eqs.~(\ref{a(t)b(t)})
and~(\ref{dalpha15}), which implies that the line
element~(\ref{metric}) becomes metric~(\ref{holographic metric}).
Therefore, the use of Eq.~(\ref{E2}) will finally give a result
consistent with those obtained by using the metric~(\ref{holographic
metric}) with Eqs.~(\ref{E1}) and~(\ref{E3}). On the other hand, it
can be shown that Eq.~(\ref{E4}) is satisfied identically by the
metric~(\ref{holographic metric}), and expressions~(\ref{p1h}) and
~(\ref{p3h}) (and therefore by the obtained holographic solution).

From Eq.~(\ref{rhodede}), we can see that there exist scenarios with
holographic energy dominating over the matter component by requiring
$\frac{\left( \alpha+1 \right)  \left( 2\,\alpha+1
\right)}{{\alpha}^{2}{\it \omega_1}+{\alpha}^{2}+2\,{\it
\omega_1}\,\alpha+\alpha+1}>2$. This relation may be rewritten as $2
\alpha^2 \omega_1+4\omega_1 \alpha -\alpha+1<0$
or equivalently
\begin{eqnarray} \label{ineq DEF}
\omega_1 < \frac{\alpha-1}{2 \alpha(\alpha+2)},
\end{eqnarray}
implying that in general the parameters $\alpha$ and $\omega_1$ vary
in the ranges $\alpha > 0$ and $\omega_1 < \frac{1}{4(\sqrt{3}+2)}$,
respectively.

It is clear that for the metric~(\ref{metrica final}) the cosmic
shear is given by~(\ref{shear1}). In this ellipsoidal cosmology the
state parameters of dark energy $\omega_{1_{{DE}}}$ and
$\omega_{3_{_{DE}}}$ are not constants and its effective parameter
of state is given by
\begin{widetext}
\begin{eqnarray}
\omega_{eff_{DE}}= \frac{\alpha(\alpha+1)(\alpha+2)(4\alpha
\omega_1+2\omega_1-\alpha+1)}{\left( {\alpha}^{2}{\it
\omega_1}+{\alpha}^{2}+2\, {\it \omega_1}\,\alpha+\alpha+1 \right)
^{2}{t}^{2-\gamma} \rho_{m0}-\alpha(\alpha+1)^2(\alpha+2)},
\label{omegaeffDE}
\end{eqnarray}
while the skewness parameter takes the form
\begin{eqnarray}
\delta_{_{DE}} = \frac{\alpha(\alpha+1)(\alpha+2)(\alpha-1)(
\omega_1-1)}{\left( {\alpha}^{2}{\it \omega_1}+{\alpha}^{2}+2\, {\it
\omega_1}\,\alpha+\alpha+1 \right) ^{2}{t}^{2-\gamma}
\rho_{m0}-\alpha(\alpha+1)^2(\alpha+2)}, \label{deltaDE}
\end{eqnarray}
\end{widetext}
where $\gamma=\frac{\left( \alpha+1 \right)  \left( 2\,\alpha+1
\right)}{{\alpha}^{2}{\it \omega_1}+{\alpha}^{2}+2\,{\it
\omega_1}\,\alpha+\alpha+1}$.

Now, we shall use the best fit values given in
Ref.~\cite{Campanelli}, which can be considered reasonable ones in
observationally testing the viability of this holographic tracking
cosmology. From Eqs.~(\ref{shear1}), (\ref{omegaeffDE})
and~(\ref{deltaDE}), we have that the model parameters must take the
values $\alpha=0.98809$, $\omega_1=0.13501$, $t_0 \rho_0=1.13869$,
in order to satisfy the best fit values $\Sigma_0=-0.004$,
$\delta_{_{DE}}=-0.05$, $\omega_{eff_{DE}}=-1.32$ of
Ref.~\cite{Campanelli}. Note that the value $\alpha=0.98809$ implies
that the free dimensionless parameter in Eq.~(\ref{holodensity}) is
constrained as $1\leq c^2< 0.99986$, implying that the
bound~(\ref{zeropoint}) will be nearly saturated.

It must be remarked that the best fit values of
Ref.~\cite{Campanelli} are obtained for ellipsoidal cosmological
models with constant state parameters of the dark energy component.
So strictly speaking, we must have constant $\omega_{1_{{DE}}}$ and
$\omega_{3_{_{DE}}}$. It can be seen that these state parameters
become constant at stages when the holographic dark energy is
dominating the cosmic evolution.

Lastly, as in the previous subsection, we shall use more tighter
constraints provided by CMB data. We have that the present shear is
constrained by $\sigma/\theta \lesssim 10^{-9}$~\cite{Russell},
which implies for the metric~(\ref{metrica final}) that $1 \leq
\alpha \leq 1.000000004$, and $0 \leq \Sigma \leq 1.33333 \times
10^{-9}$. Note that Eq.~(\ref{ineq DEF}) implies that $\omega_1<6.7
\times 10^{-10}$, so values $\omega_1 <-1/3$ (for which the
expansion is accelerated in all directions) are allowed.

\begin{figure}
\includegraphics[scale=0.3]{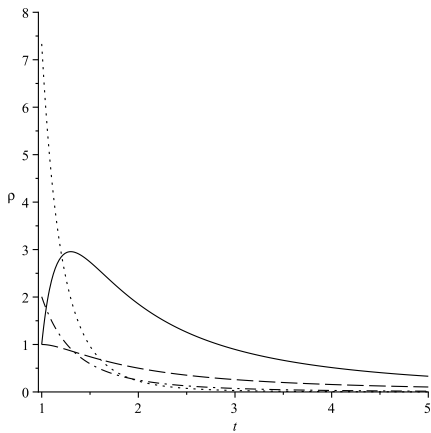}
\caption{The figure shows the qualitative behavior of energy
densities of dark matter (dotted and dash-dotted lines for
$-1<\omega<<-1/3$ and $\omega \lesssim -1/3$ respectively) and
holographic dark energy (solid and dashed lines for
$-1<\omega<<-1/3$ and $\omega \lesssim -1/3$ respectively). We see
that at $t_0$ the expansion is dominated by dark matter, and at some
$t>t_0$ the holographic dark energy begins dominate.}
\label{Fig15AA}
\end{figure}

\begin{figure}
\includegraphics[scale=0.3]{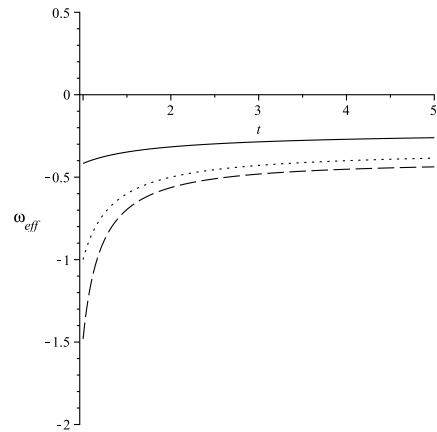}
\caption{The figure shows the qualitative behavior of effective
state parameter of holographic dark energy for $-1/3>\omega_1> 0$
(solid line), $\omega_1=-1/3$ (dotted line), and $\omega<-1/3$
(dashed line). It can be seen that for $\omega<-1/3$ at $t_0$ we
have that $\omega_{_{eff}}<-1$, so the holographic dark energy
initially behaves like phantom matter.} \label{Fig15}
\end{figure}

In Figs.~\ref{Fig15AA} and~\ref{Fig15} we show the qualitative
behavior of dark energies and effective state parameter of dark
energy for $-1<\omega_1<-1/3$. For doing this we have imposed on
dark energy the condition $\rho_{_{DE}}(t_0)=\rho_{_{DE0}}$, where
$t_0$ is a constant. Notice that for $\omega_1<-1$ the holographic
dark energy becomes negative, so we have excluded this case of our
study.

\section{Conclusion}
In this paper we have studied spatially homogeneous and anisotropic
ellipsoidal models of a universe filled with an holographic dark
energy with the Hubble length as the IR cutoff. Despite the
drawbacks with this cutoff (for obtaining a well behaved EoS for the
dark energy in FRW universes) we explored their properties and
consequences in anisotropic universes, and we have shown that in the
framework of ellipsoidal cosmologies it is possible to develop
observationally testable cosmologies.


The main result consists of the derivation of the general
ellipsoidal metric induced by a holographic energy density of the
form~(\ref{holodensity}). Essentially, the dark energy
density~(\ref{holodensity}) imposes a specific relation on the
directional scale factors of the ellipsoidal metric~(\ref{metric}),
giving the spacetime~(\ref{holographic metric}). For saturated
holographic dark energy $c=1$ (or equivalently $\alpha=1$) the flat
isotropic space is obtained. It is remarkable that for $0 \leq c^2 <
1$ the obtained metric~(\ref{holographic metric}) allows to consider
anisotropic accelerated expansion in all directions, which is in
agreement with observations. This behavior is not typical for all
Bianchi type I cosmologies since often there are solutions where
simultaneously some directional scale factors expand while others
contract.

Based on the derived metric~(\ref{holographic metric}), we develop a
tracker ellipsoidal holographic cosmology, filled with a dark matter
and a dark energy components. This solution is the ellipsoidal
version of the FRW tracker solution, for which the Friedmann
equations impose that the holographic dark energy behaves like
pressureless fluid. The ellipsoidal tracker version allows to
consider cases with $\alpha \neq 1$, so in general the holographic
dark energy does not behaves like a pressureless fluid. We show that
this ellipsoidal cosmology expands in accelerated way only in two
directions, in the third direction the expansion is decelerated.

We study also accelerated cosmic regimes where the dark matter is
neglected and the holographic dark energy dominates the expansion.
Finally, we construct an exact ellipsoidal solution, filled with
dark matter and dark energy, which has the form of the derived
metric~(\ref{holographic metric}) with variable state parameters of
the holographic dark energy.

We apply to considered holographic models the constraint values on
the shear and skewness parameters, obtained by L. Campanelli et al
by using Union2 data for supernovae~\cite{Campanelli}. These
constraints characterize the deviation from the isotropy of
ellipsoidal cosmological models, and allow our holographic models to
be ruled by an equation of state, whose range belongs to
quintessence or even phantom matter, when the dark energy is
dominating the expansion. The range~(\ref{constrain alpha}), imposed
by observations on the relevant $\alpha$-parameter, implies that the
free dimensionless parameter in Eq.~(\ref{holodensity}) is
constrained as follows: $1\leq c^2< 0.99986$, then the
bound~(\ref{zeropoint}) will be nearly saturated.

By construction, for considered holographic ellipsoidal cosmologies,
the deviation from the assumed homogeneity and isotropy of the
universe on large cosmological scales remains constant during all
evolution. This means that if the bound~(\ref{zeropoint}) is nearly
saturated today, then it remains nearly saturated for all cosmic
time.


It is interesting to note that CMB data provide tighter constraints
on the anisotropy than the SNeIa data. Specifically, for Bianchi
type I models the present shear is constrained by $\sigma/\theta
\lesssim 10^{-9}$~\cite{Russell}. We also used it for constraining
models of subsections 3B and 3C.

Another aspect that deserves consideration is that the obtained
holographic anisotropic metric~(\ref{holographic metric}) is not
asymptotically FRW for $\alpha \neq 1$. Observational constraints
allow this parameter to be $\alpha \neq 1$, although $\alpha
\approxeq 1$ ($c \approxeq 1$). This implies that observations do
not exclude the possibility of having an anisotropic expansion,
characterized by the relation $a(t)=b(t)^\alpha$ for the scale
factors. In such a way, the drawbacks with the used IR cutoff
present in holographic FRW cosmologies are substantially alleviated
in ellipsoidal scenarios.

\section{Acknowledgements}
We would like to thank Patricio Mella for useful discussions. NC and
MC acknowledge the hospitality of the Physics Department of
Universidad del B\'\i o--B\'\i o and Physics Department of
Universidad de Santiago de Chile respectively, where part of this
work was done. This work was supported by CONICYT through Grant
FONDECYT N$^0$ 1140238 (NC and MC). It also was supported by
Direcci\'on de Investigaci\'on de la Universidad del B\'\i o--B\'\i
o through grants 140807 4/R and N$^0$ GI 150407/VC (MC).

\end{document}